\documentclass[aps,twocolumn,prl,superscriptaddress,showpacs,showkeys,floatfix]{revtex4-1}

\usepackage{amsbsy} 
\usepackage{graphicx}
\usepackage{amsmath}
\usepackage{amssymb}
\input{epsf}

\begin{document}
\title{Dynamical Casimir Effect in Quantum Information Processing}

\author{Giuliano Benenti}
\affiliation{CNISM and Center for Nonlinear and Complex Systems,
Universit\`a degli Studi dell'Insubria, via Valleggio 11, 22100 Como, Italy}
\affiliation{Istituto Nazionale di Fisica Nucleare, Sezione di Milano,
via Celoria 16, 20133 Milano, Italy}
\author{Antonio D'Arrigo}
\affiliation{CNR-IMM UOS Universit\`a (MATIS), Consiglio Nazionale delle Ricerche, 
Via Santa Sofia 64, 95123 Catania, Italy}
\affiliation{Dipartimento di Fisica e Astronomia,
Universit\`a degli Studi Catania, Via Santa Sofia 64, 95123 Catania, Italy}
\author{Stefano Siccardi}
\affiliation{Department of Physics, University of Milan,
via Celoria 16, 20133 Milano, Italy}
\author{Giuliano Strini}
\affiliation{Department of Physics, University of Milan,
via Celoria 16, 20133 Milano, Italy}

\begin{abstract}
We demonstrate, in the regime of ultrastrong matter-field coupling,
the strong connection between 
the dynamical Casimir effect (DCE) and the performance
of quantum information protocols. 
Our results are illustrated by means of a realistic quantum communication
channel and show that the DCE is a fundamental
limit for quantum computation and communication and that novel
schemes are required to implement ultrafast and reliable quantum gates. 
Strategies to partially counteract the DCE are also discussed. 
\end{abstract}


\maketitle

\textit{Introduction.}
The search for high-speed operations is vital 
in quantum information~\cite{qcbook,nielsen}. 
The clock time of a quantum computer, that is
the time for the execution of a quantum gate, should be much shorter
than the decoherence time scale to allow fault-tolerant quantum 
computation. Moreover, the enhancement of transmission rates in
quantum channels is crucial to widen the applicability domain
of quantum cryptography and quantum networks.
The possibility of speeding up quantum operations is nowadays
offered by circuit quantum electrodynamics (QED)~\cite{blais,wallraff},
where one can address the ultrastrong coupling regime of light-matter
interaction. In this regime, the coupling strength $g$ becomes comparable
to the resonator frequency $\omega$~\cite{bourassa,gross,mooij}. 

The dynamical Casimir effect (DCE)~\cite{moore,dodonov}, 
that is, the generation 
of photons from the vacuum due to time-dependent boundary conditions,
has deep connections~\cite{noriRMP} with other
quantum vacuum amplification mechanisms such as the
the Hawking radiation released by black holes~\cite{hawking}
and the Unruh effect for an accelerated observer~\cite{unruh}.
Recently, the DCE has been demonstrated 
experimentally in superconducting circuit QED~\cite{norinature,lahteenmaki}.
Since a rapid variation of the matter-field coupling is needed to 
implement ultrafast quantum gates, the DCE appears as a fundamental limit
to the realization of high-speed quantum information protocols.
In this context, it is useful to remark that the coupling strength 
$g\propto 1/\sqrt{V}$, with $V$ the quantization volume for the field,
so that the DCE can be equally generated by a time-dependent coupling constant 
rather than by time-dependent boundary conditions~\cite{dodonov,dodonovJPB,dodonovPRA}. 

In this paper, we demonstrate the strong impact of photon emission 
by the DCE on quantum information processing. 
We consider a quantum channel 
for the coherent transfer of any unknown quantum state
from qubit 1 ($\textsf{Q}_1$)
to qubit 2 ($\textsf{Q}_2$), mediated by a single mode of the quantized 
electromagnetic field (cavity mode $\textsf{C}$). 
This is a genuine prototype of a \textit{quantum-bus},
which allows to reliably move quantum information and share entanglement
between different units of a quantum computing architecture.
The transmission capability of such communication channel
is quantified by its \textit{quantum capacity}~\cite{barnum,devetak},
and therefore by the channel coherent information~\cite{schumacher},
which critically depends on the coupling strength $g$.
While the system allows for a perfect transmission
in the rotating-wave approximation (RWA)
(i.e., in the limit $g/\omega\ll 1$), in the ultrastrong coupling regime
terms beyond the RWA lead to the generation of 
photons~\cite{sorbaPRB,sorbaNature}, thus spoiling the 
channel ability to convey quantum information,
up to prevent any reliable communication for very high $g$.
On the other hand, strong coupling is needed for fast transmission. 
Therefore we use the quantum information transmission rate 
(number of reliably transmitted qubits per unit time) as a figure of merit
for the channel performance. It is remarkable that the 
transmission rate is optimized for values of $g$
belonging to the ultrastrong coupling regime.
As a proof of the strong connection between the DCE and
channel performance, we show that the mean number of emitted photons is anticorrelated 
with the coherent information transmitted down the channel. 
Finally, we discuss strategies suitable to partially counteract 
photon generation by the DCE. We should stress that the quantum channel 
discussed in this paper follows steps already experimentally 
implemented by superconducting qubits coupled through a resonant 
cavity~\cite{sillanpaa}.

\newpage

\textit{Physical model.}
The qubits-cavity dynamics is described by the
Rabi Hamiltonian~\cite{micromaser}, with switcheable couplings:
\begin{equation}
  \begin{array}{c}
{\displaystyle
H(t)=H_0+H_I(t),
}
\\
\\
{\displaystyle
H_0=-\frac{1}{2}\,\sum_{k=1}^2\omega_k \sigma_z^{(k)} +
\omega\left(a^\dagger a +\frac{1}{2}\right),
}
\\
\\
{\displaystyle
H_I(t)=\sum_{k=1}^2f_k(t)\,[\,g_k \,\sigma_+^{(k)}\,(a^\dagger+a),
+g_k^\star \sigma_-^{(k)}\,(a^\dagger+a)\,],
}
\end{array}
\label{eq:noREWAquantum}
\end{equation}
where we set $\hbar=1$,
$\sigma_i^{(k)}$ ($i=x,y,z$) are the Pauli matrices for qubit $\textsf{Q}_k$ 
($k=1,2$),
$\sigma_\pm^{(k)} = \frac{1}{2}\,(\sigma_x^{(k)}\mp i \sigma_y^{(k)})$
are the rising and lowering operators for the two-level system:
$\sigma_+^{(k)} |g\rangle_k = |e\rangle_k$,
$\sigma_+^{(k)} |e\rangle_k = 0$,
$\sigma_-^{(k)} |g\rangle_k = 0$,
$\sigma_-^{(k)} |e\rangle_k = |g\rangle_k$;
the operators $a^\dagger$ and $a$ create
and annihilate a photon:
$a^\dagger |n\rangle=\sqrt{n+1}|n+1\rangle$,
$a |n\rangle=\sqrt{n}|n-1\rangle$,
$|n\rangle$ being the Fock state with $n$ photons.
The switching on/off of the couplings is governed by the functions
$f_k(t)$, in the manner detailed below.
For simplicity's sake, we
consider the resonant case ($\omega_1=\omega_2\equiv\omega$) and
the coupling strengths $g_1=g_2\equiv g\in\mathbb{R}$.
The RWA is obtained when we neglect the terms
$\sigma_+^{(k)} a^\dagger$, which simultaneously
excites $\textsf{Q}_k$ and creates a photon,
and $\sigma_-^{(k)} a$, which de-excites $\textsf{Q}_k$ and
annihilates a photon. In this limit, Hamiltonian
(\ref{eq:noREWAquantum}) reduces to the Jaynes-Cummings
Hamiltonian \cite{micromaser}, with a switchable coupling.
We set $\omega=1$, so that in the RWA the interaction time needed to transfer 
an excitation from one qubit to the field or vice versa 
($|e\rangle_k |0\rangle\leftrightarrow |g\rangle_k |1\rangle$)
is $\tau=\pi/2g$ and the vacuum Rabi frequency $\Omega_0=g$.
We work in the interaction picture, where
the effective Hamiltonian at resonance is given by
$\tilde{{H}}(t)=
e^{i{H}_0t} H_I(t) e^{-i{H}_0 t}$
(we will
omit the tilde from now on).

\begin{figure}[ht]
\includegraphics[angle=0.0, width=8.5cm]{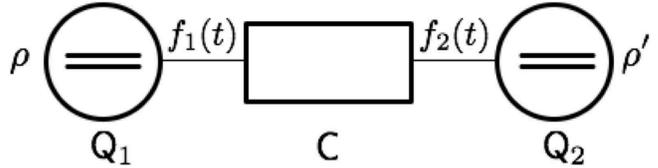}
\caption{Schematic drawing of the quantum protocol discussed in the text.
The coupling between the qubit $\textsf{Q}_i$ and the cavity $\textsf{C}$
is modulated by the function $f_i(t)$. By initially preparing
$\textsf{Q}_1$ in the state $\rho$, $\textsf{Q}_2$ is found
in the state $\rho'$ at the end of the protocol.}
\label{fig:protocol}
\end{figure}

\textit{Basic quantum protocol.}
In order to transmit quantum information through the above physical system, 
we consider the communication protocol ${\cal P}_0$ 
(sketched in Fig.~\ref{fig:protocol}),
consisting of the following steps:
\begin{itemize}
\item[1.] $\textsf{Q}_1$ is prepared in an arbitrary input state $\rho$,
while $\textsf{Q}_2$ and the cavity mode $\textsf{C}$ are in their ground state;
\item[2.] $\textsf{Q}_1$ interacts with $\textsf{C}$, for
a time $T_1=\tau$; 
\item[3.] the coupling of $\textsf{Q}_1$ with $\textsf{C}$ is switched off,
and both qubits remain non-interacting for a time $T_c$;
\item[4.] $\textsf{Q}_2$ interacts with $\textsf{C}$,
for a time $T_2=\tau$;
\item[5.] the coupling of $\textsf{Q}_2$ with $\textsf{C}$ is switched off.
\end{itemize}
The final state of $\textsf{Q}_2$ is given by
\begin{equation}
\rho'={\rm Tr}_{\textsf{Q}_1\textsf{C}}
[U (\rho\otimes |0\rangle\langle 0|\otimes|g\rangle_2{}_2\langle g|)U^\dagger],
\end{equation}
with $U$ unitary time evolution operator 
for $\textsf{Q}_1 \textsf{C} \textsf{Q}_2$,
determined by the above described
quantum protocol.
We start by considering sudden switch on/off of the couplings,
i.e. $f_1(t)=1$ for $0\le t \le T_1$, $f_1(t)=0$ otherwise;
$f_2(t)=1$ for $T_1+T_c\le t \le T_1+T_c+T_2$, $f_2(t)=0$ otherwise.
Moreover, we set $T_c=0$. 

\begin{figure}[ht]
\includegraphics[angle=0.0, width=8.5cm]{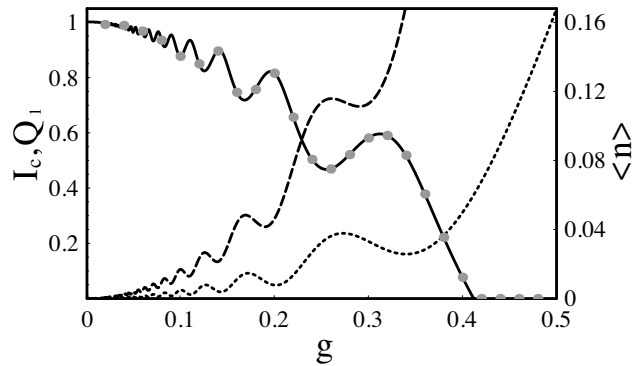}
\caption{Coherent information $I_c(\rho_u)$ (full curve, left axis),
single-shot quantum capacity $Q_1$ (gray circles,left axis)
and mean photon number $\langle n \rangle$ (right axis)
as a function of the qubit-field coupling strength $g$. 
The mean photon number is shown for 
the pure DCE (dotted curve) and at the end of the 
quantum communication protocol (dashed curve). 
The time intervals of the protocol are
$T_1=T_2=\pi/2g$ and $T_c=0$.
As we point out in the text, with a very good approximation
$Q_1\simeq I_c(\rho_u)$.
Here and in the following figures the coherent information
in computed for the maximally
mixed input state $\rho_u$.}
\label{fig:Ic}
\end{figure}

The quantum channel $\mathcal{E}$, mapping the input state $\rho$ into 
the output state $\rho'=\mathcal{E}(\rho)$,
allows for ideal quantum information
transmission, in the RWA regime and for the above suitably chosen
values of $T_1$, $T_2$, and $T_c$. 
Always in this regime, it reduces to the
amplitude damping channel for generic $T_1$, $T_2$, and $T_c$.
However, when the terms beyond
the RWA are taken into account, $\mathcal{E}$ has a non-trivial structure,
described in the supplementary material, and the channel performance
deteriorates.
The computation of the quantum capacity of channel $\mathcal{E}$,
defined as the maximum number of qubits that can be reliably transmitted
per channel use, is a 
formidable task, because one should perform an optimization over 
all possible $N$-qubit input states, for $N$ uses of the channel
and in the limit $N\to\infty$. 
Hereafter, we limit ourselves 
to the channel optimization over all possible 
single-qubit ($N=1$) input states:
\begin{equation}
\begin{array}{c}
{\displaystyle
Q_1=\max\left\{\max_\rho \left[I_c(\mathcal{E},\rho)\right],0\right\},} \\\\
{\displaystyle
I_c(\mathcal{E},\rho)\,=\,S[\mathcal{E}(\rho)]-
S_e(\mathcal{E},\rho).}
\end{array}
\end{equation}
Here the quantity $I_c$
is the \emph{coherent information}~\cite{schumacher}, 
$S(\rho)=-\mathrm{Tr}[\rho \log_2 \rho]$ the von
Neumann entropy, and 
$S_e(\mathcal{E},\rho)$ the \textit{entropy exchange}~\cite{schumacher2},
defined as
$S_e(\mathcal{E},\rho)=
S[(\mathcal{I} \otimes \mathcal{E})(|\psi\rangle\langle \psi|)]$,
where $|\psi\rangle\langle \psi|$ is any purification of $\rho$.
That is, we consider ${\textsf Q}_1$, described by the density
matrix $\rho$, as a part of a larger quantum system ${\textsf R}{\textsf Q}_1$;
$\rho=\mathrm{Tr}_{\textsf R} |\psi\rangle\langle\psi|$ and
the reference system
${\textsf R}$ evolves trivially, according to the identity
superoperator $\mathcal{I}$.
Note that, when the optimized coherent information is negative, the
single-shot quantum capacity $Q_1$ vanishes.
In the RWA limit $g\to 0$, the ideal transmission (quantum capacity
$Q=Q_1=1$) is obtained for the fully unpolarized input state 
$\rho=\rho_{u}=I/2$. We found numerically that, 
even in the ultrastrong coupling 
regime, the optimization over $\rho$ could improve 
$Q_1$ only by a tiny
amount of the order of $10^{-3}$ or smaller, with respect to
$\rho=\rho_u$. The very good agreement between $I_c(\rho_u)$ 
(full curve) and 
$Q_1$ (gray circles) is shown in Fig.~\ref{fig:Ic}. 
On this basis, in what
follows we will limit ourselves to present data for 
$I_c$ at $\rho=\rho_u$.  

\textit{Results.}
The coherent information $I_c$ as a function of the coupling strength $g$
is shown in Fig.~\ref{fig:Ic} (full curve). This quantity takes the 
value $I_c=1$, corresponding to a clean quantum channel, in the RWA limit
$g\to 0$. In the ultrastrong coupling regime ($g\gtrsim 0.1$), $I_c$ drops 
significantly and, for $g\gtrsim 0.42$, becomes
negative, so that \emph{the quantum channel can no longer be used to 
transmit quantum information}. Note that the coherent information is a non-monotonic
function of the coupling strength, with maxima at $g^{(M)}_k=1/(2k\omega+1)$
and minima at $g^{(m)}_k=1/(2k\omega)$ ($k=1,2,...$; $\omega=1$ in our units).
This regular structure, with periodicity $2\omega$ for $g^{-1}$, is a 
consequence of the terms beyond the RWA in Hamiltonian 
(\ref{eq:noREWAquantum}). Indeed, the Bloch vector (of $\textsf{Q}_1$
when $\textsf{Q}_1$ and $C$ are coupled or of $\textsf{Q}_2$
when the interaction is between $\textsf{Q}_2$ and $C$) rotates
with a speed oscillating with frequency $2\omega$ and therefore
also the distance between the exact and the RWA evolution 
exhibits oscillations of frequency $2\omega$~\cite{noRWA}.
The $2\omega$ factor can be clearly seen by expanding,
in the interaction picture, the qubit-field state at time $t$ as
$|{\Psi}(t)\rangle=\sum_{l=g,e}\sum_{n=0}^\infty
{C}_{l,n} |l,n\rangle$. 
The time-evolution of the coefficients ${C}_{l,n}$ is
governed by the equations
\begin{equation}
  \left\{
\begin{array}{l}
{\displaystyle
    i  \,\dot{C}_{g,n}(t)  =
    \Omega_n \,{C}_{e,n-1}(t) +
      \Omega_{n+1} e^{-2 i \omega t} \,C_{e,n+1}(t),
}
  \\
{\displaystyle
    i  \,\dot{C}_{e,n-1}(t)  =
    \Omega_n \,{C}_{g,n}(t) +
      \Omega_{n-1} e^{2 i \omega t} \,C_{e,n-2}(t),
}
  \end{array}\right.
  \label{eq:coefficients}
\end{equation}
with the Rabi frequencies $\Omega_n=g\sqrt{n}\,$,
where $n=0,1,2,...$ (the terms $C_{l,m}$ and $\dot{C}_{l,m}$ must be set to zero
when $m<0$).
It is interesting to remark that a decay with oscillations
in the ultrastrong coupling regime was observed for the fidelity 
of a quantum gate in Ref.~\cite{solano}. 

The strong connection between the channel performance and the DCE
is evident from the fact that
the coherent information shows a striking anticorrelation 
of peaks and valleys with the mean number  
$\langle n \rangle$ of photons in the cavity, both at the end of the protocol
(dashed curve in Fig.~\ref{fig:Ic}) and for the 
pure DCE~\cite{exotic}
(dotted curve in the same figure). In the latter case, 
qubit $\textsf{Q}_1$ and the cavity $\textsf{C}$ are prepared 
in their ground state ($\rho=\rho_g=|g\rangle_1{}_1\langle g|$)  
and the evolution of system 
$\textsf{Q}_1\textsf{C}$ is followed up to time $T_1$. Note that
the evolution of any generic input state $\rho=p|g\rangle_1{}_1\langle g|+
(1-p)|e\rangle_1{}_1\langle e|+r|g\rangle_1{}_1\langle e| +
r^\star |e\rangle_1{}_1\langle g|$ also includes the evolution of 
$\rho_g$, namely the pure DCE. 
The photons generated by the pure DCE
lead to further, stimulated emission of photons.

\textit{Strategies to contrast the DCE.} 
It is clear that suitable strategies 
must be developed to contrast the DCE in the ultrastrong coupling regime,
still allowing ultrafast quantum gates. 
Here we discuss
of two variants of the protocol ${\cal P}_0$, in which we act on the switchable
couplings $f_i(t)$.

\begin{figure}
\includegraphics[angle=0.0, width=8.5cm]{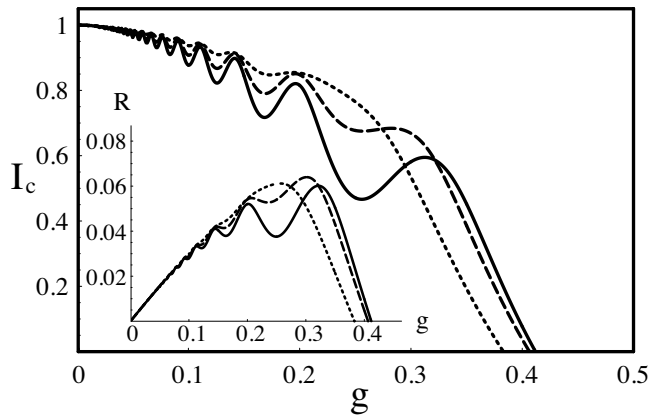}
\caption{Coherent information $I_c(\rho_u)$ (main figure) 
and transmission rate $R$ (inset) as a function
of the coupling strength $g$, for the transmission window 
discussed in the text, with $\xi=0$ (full curve), 
$\xi=0.2$ (dashed curve), and $\xi=0.5$ (dotted curve).
As in Fig.~\ref{fig:Ic}, $T_1=T_2=\pi/2g$ and $T_c=0$.} 
\label{fig:window}
\end{figure}

A first possibility (protocol ${\cal P}_1$) is to switch on/off
the interaction in a less abrupt way, for instance by substituting the 
rectangular windows for $f_1$ and $f_2$ with the Hamming window
\begin{equation}
f_k(t)=\left\{
\begin{array}{l}
1-\xi\cos(2\pi t_k/T_k)\;\;\hbox{if}\;\;0\le t_k\le T_k,
\\
0\;\;\hbox{otherwise},
\end{array}
\right.
\end{equation}
with $t_k$ time from the beginning of the window ($t_1=t$ and
$t_2=t-(T_1+T_c)$) and $0\le \xi\le 1$ ($\xi=0$ corresponds 
to the rectangular windows, $\xi=1$ to continuous functions $f_k$). 
The area below the Hamming window is the same as for the rectangular 
window, since the reduction of coupling strength at the sides of 
the window is compensated by an increase in its middle.
Such window does not affect the RWA perfect transmission, while 
relevant differences with respect to the rectangular window 
occur in the ultrastrong coupling regime. We can see 
from Fig.~\ref{fig:window} that the Hamming window leads to a significant
improvement of the coherent information 
in the region $0.1\lesssim g\lesssim 0.3$, while it can also 
deteriorate the performance of the channel at larger values of $g$. 
Moreover, the oscillations of the rectangular window instance
are smoothed.
Similar considerations can be applied to the transmission rate $R$
(see the inset of Fig.~\ref{fig:window}), defined as the ratio between
the coherent information and the duration ($T=T_1+T_c+T_2=\pi/g$) 
of the whole quantum protocol. Note that, while the coherent information 
is maximum in the RWA regime, the transmission rate is maximum
at $g\approx 0.3$. 
This individuate an \textit{optimal coupling value}
in the ultrastrong coupling regime,
which allows for the most efficient use of the physical resource
$\textsf{Q}_1 \textsf{C} \textsf{Q}_2$.

\begin{figure}
\includegraphics[angle=0.0, width=8.5cm]{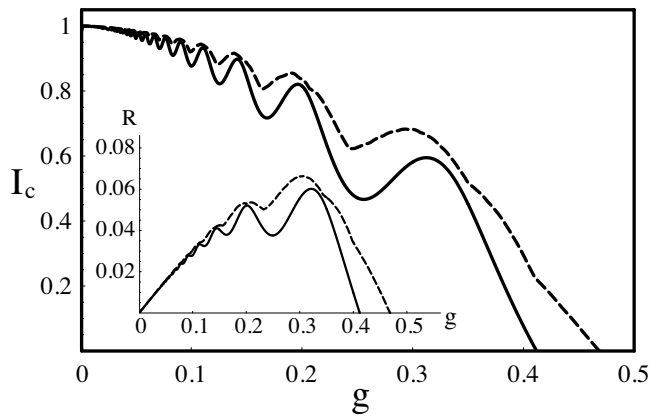}
\caption{Coherent information $I_c(\rho_u)$ (main figure) 
and transmission rate $R$ (inset) as a function
of the coupling strength $g$, for the standard protocol
(full curve) and after optimization over the times
$T_1,T_2$, and $T_c$.}
\label{fig:optimized}
\end{figure}

As a second strategy (protocol ${\cal P}_2$), 
we optimize over the timing, i.e. we optimize $I_c(\mathcal{E},\rho_u)$ 
over $T_1$, $T_2$, and $T_c$. The results of our numerical 
optimization, with the maxima of $I_c$ searched in the intervals 
$T_1,T_2\in [0.8,1.2]$ and $T_c\in [0,2\pi]$, are shown in 
Fig.~\ref{fig:optimized} (dashed curve). We can appreciate a significant 
enhancement of $I_c$ with respect to the standard timing discussed
above (full curve). Moreover, quantum information transmission 
becomes possible up to $g\approx 0.47$. Similar results are
obtained also for the optimized transmission rate, see the inset of 
Fig.~\ref{fig:optimized}.  
It is worth noting that
also for protocol ${\cal P}_2$ the optimality of the 
coupling near $g=0.3$ is confirmed.

\textit{Discussion.}
Our analysis deals with the dynamical Casimir effect (DCE) in the field of
quantum information processing,
paving the way to further investigations.
In this paper we address the aptitude 
to convey quantum information between two qubits
$\textsf{Q}_1$ and $\textsf{Q}_2$, through a cavity $\textsf{C}$.
The emergence of the DCE in the ultrastrong
coupling regime seems to put an intrinsic limit to the capability of the
bus $\textsf{Q}_1 \textsf{C} \textsf{Q}_2$ to transmit quantum information:
when $g\gtrsim 0.5$ it happens that $Q_1=0$.
Several open questions remain.
We have no evidence that the coherent information of the channel
$\textsf{Q}_1 \textsf{C} \textsf{Q}_2$ is subaddittive,
therefore one can wonder if entanglement in the input state over different
channel uses can counteract the deleterious effect of the DCE.
Moreover, after a channel use, the cavity
remains populated, as it shown in Fig.~\ref{fig:Ic} (dashed curve).
Some time has to be elapsed in order to reset the cavity
to its ground state, for instance by a suitable local control.
If instead, in order to increase the transmission rate, the 
cavity mode is not reset after 
each channel use~\cite{cavityPRL,cavityEPJD}, 
one should consider a quantum channel with 
memory effects~\cite{caruso}. 
Can memory effects be useful
in order to improve the channel performance?

While we have investigated two variants, ${\cal P}_1$ and ${\cal P}_2$
of the basic protocol ${\cal P}_0$, other improvements are surely possible,
for instance taking into account separately the various steps of ${\cal P}_0$. 
The transmission of quantum information is realized by two consecutive channels,
${\cal E}={\cal E}_2\circ{\cal E}_1$, where
${\cal E}_1:\textsf{Q}_1 \rightarrow \textsf{C}$ and 
${\cal E}_2:\textsf{C} \rightarrow \textsf{Q}_2$.
By a numerical analysis we found that the channel ${\cal E}_1$
succeeds in reliably transmitting quantum information from the 
first qubit to the cavity, also in  cases when $Q_1=0$;
for example for $g=0.5$ we have that $Q^{({\cal E}_1)}>0.75$. 
It is the further processing
of the quantum information by ${\cal E}_2$ which produces a vanishing $Q_1$.
After the first channel, the information is spread up over different levels
of the cavity as a consequence of the DCE:
this is the cause that prevents the second channel to work properly.
Modification of the basic protocol ${\cal P}_0$ 
and novel schemes could be studied 
in order to contrast the DCE, taking into account the different 
performances of quantum channels ${\cal E}_1$ and ${\cal E}_2$.
Techniques such as the  quantum optimal control~\cite{simone}
might be useful, or one could also take inspiration from
counterintuitive protocols for population transfer in stimulated 
Raman adiabatic passage (STIRAP)~\cite{STIRAP}.

To summarize, we have illustrated in the regime of ultrastrong
matter-field coupling the connection between the 
dynamical Casimir effect and the performance
of quantum information protocols.
Since the ultrastrong regime is already investigated 
in circuit QED experiments, it can be foreseen that the
DCE will play for quantum computation and communication 
a role similar to the one played by the (static) Casimir 
effect in the development of nanomechanical tecnologies~\cite{capasso}.

\begin{acknowledgments}
G.B. acknowledges the support by MIUR-PRIN project
``Collective quantum phenomena: From strongly correlated systems to
quantum simulators''.
A.D. acknowledges support from CSFNSM Catania.
\end{acknowledgments}


\section*{Supplementary material}

\begin{figure*}[t!]
  \begin{center}
  \includegraphics[width=16.0cm]{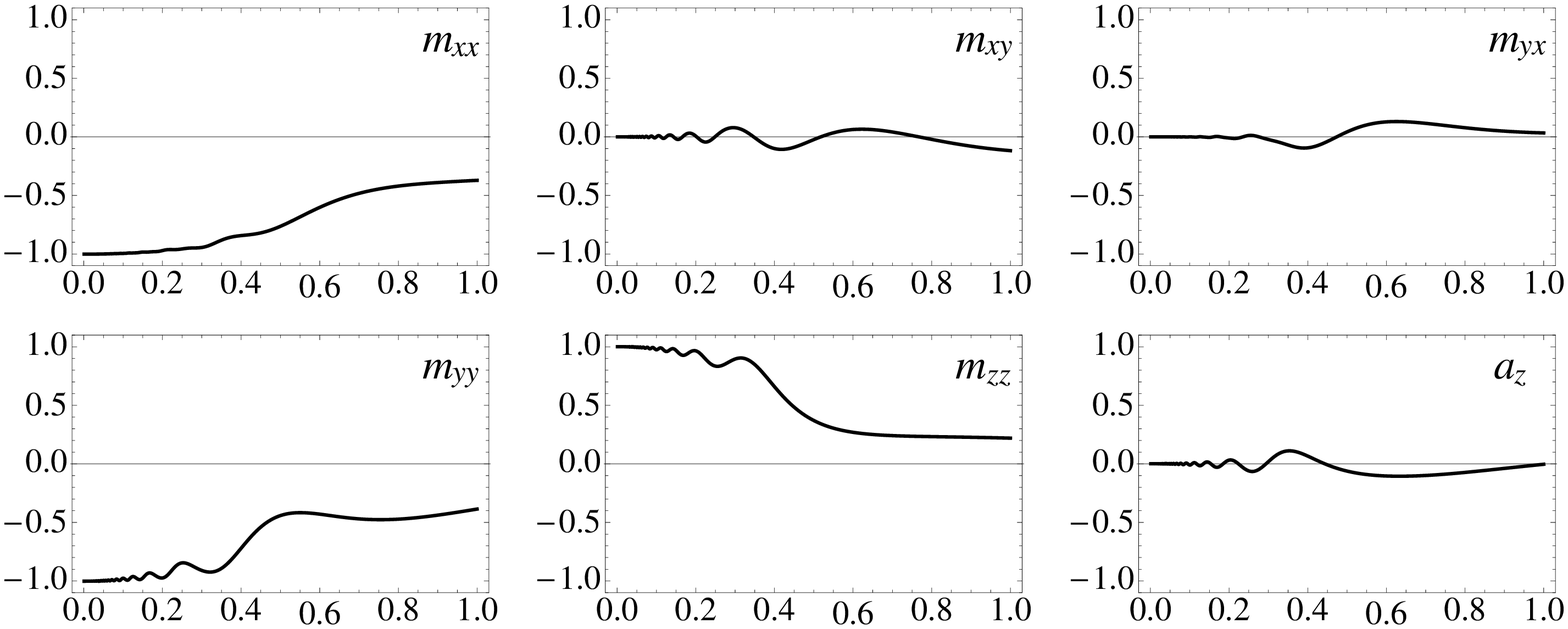}
  \end{center}
  \caption{Non-zero parameters of the Fano representation of the
quantum channel $\mathcal{E}$, as a function of the
coupling strength $g$, for $T_1=T_2=\pi/2g$, $T_c=0$, and sudden switch
on/off of the couplings.}
  \label{fig:blochg}
\end{figure*}

The quantum channel $\mathcal{E}$ introduced in the main text,
mapping the input state
$\rho$ into the output state $\rho'$, namely
$\rho'=\mathcal{E}(\rho)$, can be conveniently described
in the Fano representation (also known as the Bloch representation)
~\cite{fano,eberly,mahler,striniqpt,strinidiamondnorm}.
In the RWA regime, the quantum protocol described by $\mathcal{E}$, 
transfers, up to a trivial unitary transformation,
the state $\rho$ from $\textsf{Q}_1$ to the cavity $\textsf{C}$,
and finally from $\textsf{C}$ to $\textsf{Q}_2$,
leaving $\textsf{Q}_1$ and $\textsf{C}$ in their ground state.
More precisely, if ${\bf r}^\prime=(x',y',z')$ are the Bloch ball coordinates
of the final state $\rho^\prime$ of $\textsf{Q}_2$ and ${\bf r}=(x,y,z)$ the
coordinates of the input state $\rho$ of $\textsf{Q}_1$, then
$x'=-x$, $y'=-y$, and $z'=z$. The state $\rho$ can therefore be
recovered from $\rho'$ after a rotation of angle $\pi$ about
the $z$-axis of its Bloch ball. Deviations from the ideal
quantum protocol appear when effects beyond the RWA cannot be neglected.
In the Fano form we write
$\rho=\frac{1}{2}\left(I^{(1)}+{\bf r}\cdot {\bf \sigma}^{(1)}\right)$ and
$\rho^\prime=\frac{1}{2}\left(I^{(2)}+{\bf r}^\prime\cdot {\bf \sigma}^{(2)}\right)$, 
with $I^{(k)}$ identity operator for qubit $k$.
Due to the linearity of quantum mechanics the Bloch vectors
${\bf r}$ and ${\bf r'}$ are connected through an affine map $\mathcal{M}$
as follows:
\begin{equation}
\left[
\begin{array}{c}
{\bf r'}
\\
\hline
1
\end{array}
\right]
=
\mathcal{M}
\left[
\begin{array}{c}
{\bf r}
\\
\hline
1
\end{array}
\right]
=
\left[
\begin{array}{ccc}
  {\bf M} & \Big \lvert & {\bf a}  \\
 \hline
  {\bf 0}^T & \Big \lvert & 1
\end{array}
\right]
\left[
\begin{array}{c}
{\bf r}
\\
\hline
1
\end{array}
\right],
\label{eq:affine}
\end{equation}
where ${\bf M}$ is a $3\times 3$ real matrix,
${\bf r}$, ${\bf r'}$ and ${\bf a}$ real column vectors of
dimension $3$ and
${\bf 0}$ the null vector of the same dimension.
The Fano representation of quantum operations is physically transparent
since the Bloch vectors directly provide the expectation values
of polarization measurements.
While in general an affine map for a qubit depends
on twelve parameter~\cite{qcbook},
we found from the numerical simulation of the above described
quantum protocol the following structure
of ${\bf M}$ and ${\bf a}$:
\begin{equation}
{\bf M} = \left(
\begin{array}{ccc}
  m_{xx} & m_{xy} & 0        \\
  m_{yx} & m_{yy} & 0        \\
    0    &   0    & m_{zz}   \\
  \end{array} \right), \qquad
{\bf a} = \left(
\begin{array}{c}
    0   \\
    0   \\
    a_z \\
  \end{array} \right).
\label{eq:Memoryless-Kraus-Operators}
\end{equation}
The dependence of the six non-zero parameters,
$m_{xx},m_{xy},m_{yx},m_{yy},m_{zz}$, and $a_z$,
is shown in Fig.~\ref{fig:blochg} as a function
of the parameter $g$. Note that in the RWA
limit ($g\ll 1$) we have $m_{xx}=m_{yy}=-1$,
$m_{zz}=1$, and $m_{xy}=m_{yx}=a_z=0$, as expected
for the ideal quantum state transfer protocol.
On the other hand, for $g\gtrsim 0.1$ significant
deviations from the ideal protocol are observed.
It is interesting to remark that the positions of peaks and valleys
matches those found in Fig.~2 of the main text for the coherent 
information (note that here values of $g$ up to $g=1$ are considered).
Quantum channel $\mathcal{E}$ has an interesting
and non trivial
structure, since it is nonunital ($\mathcal{E}
(I)\ne I$ since $a_z\ne 0$) and matrix $M$ is not
symmetric ($m_{xy}\ne m_{yx}$).

The geometrical meaning of the quantum channel $\mathcal{E}$ can
be understood from the decomposition of map
(\ref{eq:affine})-(\ref{eq:Memoryless-Kraus-Operators}) into
a sequence of elementary affine maps.
We first write $\mathcal{M}=\mathcal{M}_1\mathcal{M'}$,
where
\begin{equation}
\mathcal{M}_1=\left[
\begin{array}{cccc}
\cos\theta & 0 & 0 & 0 \\
0 & \cos\theta & 0 & 0 \\
0 & 0 & \cos^2\theta & \pm \sin^2\theta \\
0 & 0 & 0 & 1
\end{array}
\right],
\end{equation}
represents a displacements of the Bloch sphere along the
$z$-axis~\cite{striniepjd}.
Note that we have $\sin^2\theta=a_z$ or
$\sin^2\theta=-a_z$ depending on the sign of $a_z$.
In the first case the displacement of the center of the Bloch sphere is
along the positive direction of the $z$-axis and can be seen as representative
of zero temperature dissipation (amplidude damping
channel~\cite{qcbook}), in the latter case
the displacement is along the negative $z$-direction and can be seen as
thermal excitation.
The affine map $\mathcal{M}'$ represents a unital quantum channel. It reads as follows:
\begin{equation}
\mathcal{M'}=
\left[
\begin{array}{ccc}
  {\bf M'} & \Big \lvert & {\bf 0}  \\
 \hline
  {\bf 0}^T & \Big \lvert & 1
\end{array}
\right]
=
\left[
\begin{array}{cccc}
m_{xx}' & m_{xy}' & 0 & 0 \\
m_{yx}' & m_{yy}' & 0 & 0 \\
0 & 0 &  m_{zz}'& 0 \\
0 & 0 & 0 & 1
\end{array}
\right],
\end{equation}
with $m_{ij}'=(\cos\theta) m_{ij}$ for $i,j=x,y$ and
$m_{zz}'=(\cos^2\theta) m_{zz}$.

Matrix ${\bf M'}$ can be written using the singular value
decomposition as
${\bf M'}={\bf O_1 D O_2^T}$,
with ${\bf O_1}$ and ${\bf O_2}$ rotation matrices and
${\bf D}$ diagonal scaling matrix.
Since ${\bf S}\equiv {\bf M' M'^T}=
{\bf O_1 D^2 O_1^T}$, the diagonal entries of
of ${\bf D}$ (known as the singular values) are the
square roots of the eigenvalues of the symmetric matrix ${\bf S}$.
We can therefore write the affine map
${\cal M'}$ as composition of three elementary maps,
${\cal M'}={\cal M}_2{\cal M}_3{\cal M}_4$,
with ${\cal M}_2$ and ${\cal M}_4$ rotations of the Bloch sphere
about the $z$ axis
and ${\cal M}_3$ deformation of the Bloch sphere into an ellipsoid
centred at the origin of the Bloch sphere and whose axes are directed
along $x$, $y$ and $z$. The lengths of the semi-axes of the ellipsoid
are the singular values of ${\bf M'}$.

To summarize, the overall quantum channel $\mathcal{E}$
is obtained by the composition of
a rotation of the Bloch sphere 
(affine map $\mathcal{M}_4$), a deformation
of the Bloch sphere ($\mathcal{M}_3$), another rotation 
($\mathcal{M}_2$), and a displacement of the Bloch sphere
($\mathcal{M}_1$). We need three real parameters to determine
$\mathcal{M}_3$, and one parameter for each of the other
transformations. Overall we have six real parameters, as also
clear from Eq.~(\ref{eq:Memoryless-Kraus-Operators}).

Finally, we point out that a set of Kraus operators for map
$\mathcal{E}$ can be easily obtained by composing the
Kraus representations for the elementary transformations
$\mathcal{M}_i$, ($i=1,...,4$), for which Kraus operators
are well known~\cite{qcbook}.


\end{document}